# Deep Learning *GW* Quasiparticle Hamiltonians for Many-Body Excited-State Electronic Structure at Scale

Xiaoxun Gong,[1,2] Zechen Tang,[3] Woochang Kim,[1,2] Yang Li,[3] Wenhui Duan,[3,4] Yong Xu[3] and Steven G. Louie[1,2,*]

[1] Department of Physics, University of California at Berkeley, Berkeley, California 94720, USA

[2] Materials Sciences Division, Lawrence Berkeley National Laboratory, Berkeley, California 94720, USA

[3] Department of Physics, Tsinghua University, Beijing 100084, China

[4] Chen-Ning Yang Institute for Advanced Study, Tsinghua University, Beijing 100084, China

[*] sglouie@berkeley.edu

Accurate quasiparticle electronic structures are the foundation for understanding excited-state properties of materials and explaining optoelectronic, quantum, and transport phenomena. First-principles *GW* calculations nevertheless remain computationally intensive for large or configurationally complex systems. Here we introduce DeepH-*GW*, a deep-learning framework that predicts an effective *GW* quasiparticle Hamiltonian directly from atomic structure. Building on the local, equivariant message-passing architecture of DeepH, DeepH-*GW* is trained on high-fidelity plane-wave *GW* calculations through a real-space Hamiltonian-reconstruction interface. This approach combines the systematic accuracy and broad chemical applicability of plane-wave methods with linear-scaling neural-network inference. Across the systems examined, DeepH-*GW* reproduces quasiparticle band structures with errors on the order of a few meV. Moreover, we show that, despite the intrinsic nonlocality of many-body interactions, DeepH-*GW* exhibits strong cross-scale transferability from relatively small training structures to substantially larger supercells. We demonstrate that the framework can accurately capture *GW*-level electron-phonon band-gap renormalization through predictions in supercells with thermal displacements, illustrating the power of the approach. DeepH-*GW* therefore provides a practical route toward large-scale many-body simulations and foundation models for excited-state electronic structure.

Reliable modeling of many-body excited-state phenomena is essential for understanding and designing quantum materials, semiconductor devices, optoelectronic systems, and other systems whose functionalities are governed by interactions and excitations. Although standard density functional theory (DFT) provides an efficient framework for ground-state properties, its Kohn-Sham eigenvalues generally cannot be interpreted as a quantitatively reliable description of quasiparticles [1]. Many-body perturbation theory within the *GW* approximation addresses this limitation by solving a Dyson equation with a nonlocal, frequency-dependent self-energy, and has consequently become a standard first-principles approach for predictive studies of electronic structure in the excited-state regime for many materials [2–7]. Its computational cost and demanding convergence requirements, however, limit its application to large structures, complex atomic configurations, and high-throughput materials discovery. This limitation exemplifies a fundamental accuracy-efficiency dilemma: because higher fidelity comes at the cost of increased computational effort, the structurally complex materials or broad configurational spaces for which precise excited-state predictions are most valuable are often precisely those least accessible to conventional *GW* calculations.

Deep learning offers a promising route through this dilemma. Data-driven models have already accelerated the prediction of potential energy surfaces, charge densities, wavefunctions, and electronic Hamiltonians at the DFT level [8–10]. In particular, the DeepH framework learns the mapping from local atomic structure to the DFT Hamiltonian under localized atomic-orbital (AO) bases using equivariant message-passing neural networks [11–15]. By exploiting nearsightedness of electronic matter and enforcing fundamental spatial symmetries, DFT-based DeepH models trained on relatively small structures can be transferred to much larger systems while retaining high accuracy, thereby enabling efficient electronic-structure calculations at scale.

Extending this paradigm from the DFT level to the quasiparticle level of many-body perturbation theory is appealing but nontrivial. First, the long-range many-electron interactions (the self-energy operator) underlying the *GW* formalism appear, at first sight, to conflict with the locality required by scalable message-passing models. Second, generating high-fidelity *GW* training data is computationally demanding and involves multiple numerical subtleties [3,6,7]. Therefore, significant challenges remain in developing a successful deep-learning *GW* framework that retains the accuracy needed to resolve subtle quasiparticle-energy variations while remaining transferable across system sizes.

Several studies have explored combining machine learning with *GW* and related many-body methods. One class of approaches directly predicts *GW*-level observables, including quasiparticle energies, photoemission spectra, and core-electron binding energies, from atomic structures and DFT-derived descriptors [16–23]. A second class learns frequency-dependent self-energies or Green's functions, from which quasiparticle and spectral properties can be obtained through Dyson's equation [24–26]. A third class focuses on dielectric screening or density-response functions, thereby accelerating key intermediate stages of *GW* or *GW*-Bethe-Salpeter calculations [27,28]. These studies have firmly indicated that machine learning can capture meaningful many-body corrections and substantially reduce computational expense. Nevertheless, many existing methods focus on selected observables, achieve limited accuracy or transferability, or have been demonstrated primarily for molecules or relatively small material systems. A transferable framework that predicts *GW*-level electronic structure directly from atomic structure, while maintaining efficiency and accuracy when applied to substantially larger periodic systems, has yet to be established.

Here we present DeepH-*GW*, a deep-learning framework that predicts an effective *GW* quasiparticle Hamiltonian directly from atomic structure. By interfacing the local, symmetry-aware DeepH architecture with plane-wave (PW) *GW* calculations through Hamiltonian reconstruction, DeepH-*GW* combines high-fidelity reference calculations with efficient Hamiltonian prediction. Despite the nonlocal character of the *GW* self-energy, the resulting models exhibit excellent cross-scale transferability and accurately predict quasiparticle electronic structures for systems substantially larger than those included in training. Neural network inference scales linearly with system size and yields quasiparticle energies over a range of 10 eV from the Fermi level, with errors on the order of a few meV, in the systems considered here. We further demonstrate the ability of DeepH-*GW* to predict quasiparticle energies in phonon-perturbed supercells and thereby evaluate the temperature-dependent electron-phonon band gap renormalization of semiconductors, which would otherwise require prohibitively expensive repeated *GW* calculations. These capabilities open opportunities for many-body simulations of large-sized structures such as moiré supercells, quantum defects, junctions, and heterogeneous materials containing thousands of atoms. More broadly, DeepH-*GW* establishes a pathway toward *GW*-level electronic-structure foundation models and autonomous discovery workflows for materials with engineered excitonic, optoelectronic, and quantum functionalities.

**Theory**

The central challenge in constructing an accurate and scalable framework for learning *GW*-level quasiparticle properties is the construction of a physically meaningful and reusable many-body electronic-structure representation that bridges deep-learning and the many-body perturbation theory formalism of the *GW* method. Here, starting with accurate *GW* results (e.g., from converged PW calculations in the training set), we construct an effective *GW* Hamiltonian expressed under a localized AO basis set. The neural network learns the mapping from the local atomic environment to this Hamiltonian, as illustrated in Fig. 1:

$$H_{i\alpha,j\beta}^{GW} = \left\langle \phi_{i\alpha} \middle| \frac{\hat{p}^2}{2m} + \hat{V}_{\text{ion}} + \hat{V}_{\text{Har}} \middle| \phi_{j\beta} \right\rangle + \sum_{nm\boldsymbol{k}} \langle \phi_{i\alpha} | \psi_{n\boldsymbol{k}} \rangle \, \Sigma_{nm\boldsymbol{k}}^{\text{eff}} \langle \psi_{m\boldsymbol{k}} | \phi_{j\beta} \rangle. \tag{1}$$

Here, $|\phi_{i\alpha}\rangle$ denotes an AO basis function centered at atom $i$ with orbital index $\alpha$. The operators $\frac{\hat{p}^2}{2m}$, $\hat{V}_{\text{ion}}$ and $\hat{V}_{\text{Har}}$ are the kinetic energy, ionic potential, and Hartree potential operators, respectively. $|\psi_{n\boldsymbol{k}}\rangle$ is the quasiparticle wavefunction of band $n$ at wavevector $\boldsymbol{k}$, normalized in the Born-von Kármán supercell. $\Sigma_{nm\boldsymbol{k}}^{\text{eff}}$ is a static, Hermitian effective self-energy of the form commonly used in the quasiparticle self-consistent *GW* (QS*GW*) approach [29,30]:

$$\Sigma_{nm\boldsymbol{k}}^{\text{eff}} = \frac{1}{4}\left(\Sigma_{nm\boldsymbol{k}}(\varepsilon_{n\boldsymbol{k}}) + \Sigma_{mn\boldsymbol{k}}^{*}(\varepsilon_{n\boldsymbol{k}}) + \Sigma_{nm\boldsymbol{k}}(\varepsilon_{m\boldsymbol{k}}) + \Sigma_{mn\boldsymbol{k}}^{*}(\varepsilon_{m\boldsymbol{k}})\right), \tag{2}$$

where

$$\Sigma_{nm\boldsymbol{k}}(E) = \langle \psi_{n\boldsymbol{k}} | \hat{\Sigma}(E) | \psi_{m\boldsymbol{k}} \rangle. \tag{3}$$

Once this effective Hamiltonian is obtained, its diagonalization yields the quasiparticle energies and wavefunctions. These quantities can subsequently be used to evaluate excited-state observables and response functions of the system. In the following, we discuss several conceptual/physical and computational challenges addressed by this construction, together with the approximations and implementation choices adopted in the present work.

*Locality and nearsightedness in real space*. — The first challenge is resolving the conflict between the intrinsic long-range interactions and nonlocal correlations entering the *GW* formalism and the locality and nearsightedness required for linear scaling and transfer across system sizes. Although related, locality and nearsightedness are two distinct properties. Locality requires that the target of learning must be spatially local, or at least numerically short ranged, meaning that a matrix element decays sufficiently fast when its two spatial arguments move apart. Nearsightedness requires that a local object will be negligibly affected when the change of atomic environment is sufficiently far away.

The Kohn-Sham Hamiltonian within semi-local approximations for the exchange-correlation potential in DFT is a local object, whereas the *GW* self-energy is generally nonlocal. Nevertheless, Hamann and Vanderbilt [31] showed that QS*GW* quasiparticle states admit well-localized Wannier representations and that the corresponding static quasiparticle Hamiltonian can be accurately interpolated using real-space Wannier matrix elements. Wannier interpolation has thus far become a standard approach for evaluating *GW* band structures on a fine $\boldsymbol{k}$-mesh or $\boldsymbol{k}$-path. A recent study by Dernek et al. [32] further provided numerical evidence that the static, Hermitian QS*GW* effective self-energy is sufficiently short-ranged. These results support the use of the effective *GW* Hamiltonian in Eq. (1) as a local learning target within a message-passing framework.

The nearsightedness of electronic quantities, especially the DFT Hamiltonian and density matrix, has been extensively studied [33–37]. The nearsightedness of the Hamiltonian also underlies the real-space interpolation of electron-phonon matrix elements using Wannier functions, which is now routinely used when dense sampling of electronic and phonon wavevectors is needed for modeling of electron-phonon phenomena [34,38,39]. Important caveats are well-known to researchers. Long-range electrostatic interactions arise in polar insulators and in heterostructures with charge transfer, and metals can exhibit slow-decaying Friedel oscillations. Local machine-learning methods must therefore be applied carefully to such systems or augmented with explicit long-range contributions.

The nearsightedness of the *GW* self-energy is less extensively studied, but recent calculations have shown that electron-phonon matrix elements evaluated within the *GW* perturbation theory (*GW*PT) can also be accurately interpolated using Wannier functions, provided that the long-range Fröhlich interaction in polar materials is handled separately [40–42]. This provides further numerical evidence that the AO effective Hamiltonian remains suitable for local machine learning models.

*AO and PW formulations of GW*. — The majority of previously developed Hamiltonian-learning methods use reference data generated by electronic-structure codes based on localized AO basis sets, including those for *GW* calculations. Localized bases are computationally efficient and particularly suitable for molecular systems. Because they are localized around atoms by construction, they naturally provide the real-space structure needed by local machine-learning frameworks. In contrast, PWs are extended over space and do not provide such a local structure.

However, PW *GW* calculations offer systematic control over basis-set convergence and are particularly well suited to chemically diverse periodic materials, making them attractive for generating consistent, high-fidelity datasets spanning broad configurational and chemical spaces. Moreover, converged *GW* calculations often require a large number of high-energy unoccupied states, which may be difficult to represent accurately with a limited number of localized AO bases. An effective deep-learning framework at the *GW* level should therefore bridge the gap between the PW and AO approaches, combining localized Hamiltonian learning with the accuracy, flexibility, and systematically controllable convergence of PW calculations.

A recently developed real-space reconstruction method connects DeepH with PW DFT calculations [12]. Equation (1) is a natural generalization of this method to the *GW* level. In Eq. (1), the Hartree potential, self-energy matrix elements, and quasiparticle wavefunctions are obtained from converged PW calculations, while the resulting quasiparticle Hamiltonian is reconstructed in a localized AO basis using the formalism of Ref. [12]. The errors due to reconstruction can be reduced systematically by improving the AO basis and increasing the number of bands included in the reconstruction.

Importantly, the AO basis for representing this reconstructed Hamiltonian need not reproduce the entire quasiparticle spectrum with uniform accuracy. A relatively compact basis is sufficient when downstream applications involve only states within a specified energy range around the Fermi level. This restriction substantially reduces the cost of constructing the training targets, training the neural network, evaluating the predicted Hamiltonian, and further diagonalizing it for quasiparticle energies and wavefunctions.

*Controlling the computational cost.* — Generating high-fidelity training data at the *GW* level is considerably more difficult than at the DFT level. *GW* calculations are computationally demanding and involve multiple numerical subtleties [3,6]. Better convergence demands more computational effort, whereas insufficient convergence not only adds noise to a training dataset but can also yield qualitatively incorrect or unphysical quasiparticle properties. Several physically motivated approximations can mitigate this cost. In this work, all our *GW* calculations are performed at the one-shot $G_0W_0$ level. We neglect all off-diagonal matrix elements of $\hat{\Sigma} - \hat{V}_{\mathrm{xc}}$, where $\hat{V}_{\mathrm{xc}}$ is the DFT exchange-correlation potential. Furthermore, we only apply the *GW* correction within a limited window around the Fermi level. Under these approximations, Eq. (1) reduces to

$$H^{G_0W_0}_{i\alpha,j\beta} = H^{\mathrm{DFT}}_{i\alpha,j\beta} + \sum_{n=n_{\min}}^{n_{\max}} \sum_{\boldsymbol{k}} f(\varepsilon^{\mathrm{DFT}}_{n\boldsymbol{k}}) \langle \phi_{i\alpha} | \psi^{\mathrm{DFT}}_{n\boldsymbol{k}} \rangle \left( \varepsilon^{G_0W_0}_{n\boldsymbol{k}} - \varepsilon^{\mathrm{DFT}}_{n\boldsymbol{k}} \right) \langle \psi^{\mathrm{DFT}}_{n\boldsymbol{k}} | \phi_{j\beta} \rangle, \qquad (4)$$

where $H^{\mathrm{DFT}}_{i\alpha,j\beta}$ is the AO DFT Hamiltonian reconstructed from PW DFT calculations using the real-space reconstruction method [12], $\varepsilon^{\mathrm{DFT}}_{n\boldsymbol{k}}$ is the Kohn-Sham band energy, $\varepsilon^{G_0W_0}_{n\boldsymbol{k}}$ is the one-shot $G_0W_0$ quasiparticle energy, and $|\psi^{\mathrm{DFT}}_{n\boldsymbol{k}}\rangle$ is the Kohn-Sham wavefunction. The interval $[n_{\min}, n_{\max}]$ defines the set of bands for which *GW* quasiparticle energies are used, and $f$ is a window function [43].

This simplification from QS*GW* to $G_0W_0$ does not fundamentally invalidate the locality and nearsightedness considerations discussed above. Self-consistency quantitatively modifies the quasiparticle energies and wavefunctions, but it does not introduce any new mechanism governing locality or nearsightedness. Locality of the Hamiltonian in real space comes from the smoothness of the eigenvalues and wavefunctions in $\boldsymbol{k}$ space, and its nearsightedness is primarily associated with electronic screening. In practice, Wannier interpolation is routinely applied to quasiparticle energies at the $G_0W_0$ level [44]. Similarly, *GW*PT matrix elements computed within a one-shot $G_0W_0$ framework can be Wannier-interpolated [40,42,45]. We therefore expect the Hamiltonian constructed by Eq. (4) to preserve similar locality and nearsightedness required by DeepH, subject to the material-specific caveats discussed above.

*Frequency dependence of the self-energy.* — The frequency dependence of the dynamical *GW* self-energy makes it fundamentally different from the static DFT exchange-correlation potential. Equations (1) and (4) eliminate this explicit frequency dependence by constructing an effective static quasiparticle Hamiltonian. There are other possible choices, but each introduces additional complications.

Directly learning the frequency-dependent self-energy on the real-frequency axis is particularly challenging because it exhibits strong pole-induced variations with frequency, and the self-energy at a specific real frequency is not guaranteed to be local [7,46]. On the imaginary-frequency axis, the self-energy is considerably smoother, motivating previous studies that use it as the learning target [24–26]. However, most quasiparticle observables require information about the self-energy on the real frequency axis, so a subsequent analytic continuation of the self-energy is required to recover its real-frequency behavior. Although it has proven to be successful for a series of material systems, this step is numerically ill-conditioned, and can introduce errors that depend sensitively on the assumed continuation form [46]. A related possibility is to learn the imaginary-time self-energy $\Sigma(i\tau)$. For insulating systems, this quantity is localized in real space. This has motivated the *GW* space-time method for efficient *GW* calculations [47,48]. Nevertheless, extracting real-frequency quasiparticle properties again requires analytic continuation. Compared with these possible approaches, for the applications considered here, learning a static effective quasiparticle Hamiltonian is therefore the most direct and numerically robust.

*Deep-learning architecture*. — Apart from replacing the AO DFT Hamiltonian with an AO effective *GW* Hamiltonian as the learning target, we retain the established DeepH neural-network architecture [11,14,15]. The effective *GW* Hamiltonian transforms under the Euclidean group (E(3)) in the same manner as the DFT Hamiltonian. The existing E(3)-equivariant architecture can therefore be applied without modification to its symmetry constraints.

A common strategy in multi-fidelity machine learning is to use outputs from a lower-level electronic-structure method as descriptors for predicting results from a higher-level method. Such descriptors provide information beyond the structural geometry at a comparatively modest additional cost. Previous machine-learning models for *GW*-level properties have used DFT or Hartree-Fock quantities derived from DFT wavefunctions [21–23], as well as Hamiltonians, density matrices, mean-field Green's functions [17,24,26], and other types of derived features. These studies generally find that lower-level electronic descriptors improve the accuracy of the predicted many-body level quantities.

We do not adopt this strategy in the present work, primarily to preserve the simplicity and scalability of the formalism. In principle, the mapping from atomic structure to the AO effective *GW* Hamiltonian is well defined and does not require additional electronic-structure inputs. Moreover, the inclusion of conventional DFT-derived descriptors would generally remove the linear-scaling advantage of the inference workflow unless an order-$N$ DFT method were used [49]. For the material systems studied here, a model with only atomic structure as input already achieves satisfactory accuracy. Nevertheless, incorporating inexpensive DFT-level descriptors is a promising direction for future development.

## Results

DeepH-*GW* combines four capabilities that are challenging to achieve simultaneously but needed: high-fidelity PW *GW* reference data, a localized and symmetry-preserving Hamiltonian representation, linearly scaling neural-network inference, and transfer from relatively small training structures to much larger supercells. We first validate the reconstruction of the AO effective quasiparticle Hamiltonian from PW calculations. We then assess the accuracy of neural-network predictions for structures comparable in size to those used in training and subsequently for substantially larger supercells. Finally, we demonstrate that the resulting scale transfer enables a practical calculation of $GW$-level electron-phonon band-gap renormalization using a formalism with supercells containing thousands of atoms.

The practical workflow of building the AO effective $GW$ Hamiltonian is illustrated in Fig. 1(b). PW DFT calculations are performed using the Quantum ESPRESSO package [50] with norm-conserving pseudopotentials [51], and the outputs provide the starting point for subsequent PW $GW$ calculations. AO basis sets are generated using the SIESTA package [49] with the same set of pseudopotentials. $GW$ calculations are performed under the PW formalism using the BerkeleyGW package [52]. The resulting DFT and $GW$ quantities are then collected, and the AO effective $GW$ Hamiltonian is reconstructed with the HPRO package [12] according to Eq. (4).

*Validation of Hamiltonian reconstruction*. — We benchmark the DeepH-$GW$ framework on two representative semiconductors: diamond and GaP. We first validate the reconstruction process from the PW representation to the AO reconstructed Hamiltonian. At the DFT level, this test establishes whether the selected AO basis accurately spans the relevant PW Kohn-Sham states. At the $GW$ level, it determines whether the effective Hamiltonian defined by Eq. (4) faithfully reproduces the quasiparticle energies obtained from the original PW calculation.

As shown in Fig. 2, the band structures obtained from the reconstructed AO DFT Hamiltonians accurately reproduce the PW DFT bands for the unit cells of both diamond and GaP, confirming that the selected AO basis sets accurately describe the relevant Kohn–Sham states. At the $GW$ level, diagonalization of the reconstructed effective Hamiltonian also yields good overall agreement with the PW $G_0W_0$ band structures, although the deviations are somewhat larger than at the DFT level, particularly for some conduction bands. These discrepancies may arise from the finite AO basis, the spatially nonlocal character of the $GW$ self-energy, and the truncation of the corrected band manifold in Eq. (4). The reconstruction accuracy can be improved systematically by optimizing the AO basis and increasing the number of bands included in the quasiparticle correction. The parameters used here represent a balance between accuracy and computational cost and are sufficient for evaluating the subsequent machine-learning step.

*Accuracy and scale transfer of DeepH-$GW$ models*. — We next train DeepH-$GW$ models to predict the reconstructed effective AO Hamiltonians. Separate datasets of $3 \times 3 \times 3$ supercells with random atomic displacements were generated for diamond (maximum 0.1 Å) and GaP (maximum 0.15 Å). After training one model for each material, we first evaluate the models on randomly displaced $3 \times 3 \times 3$ supercells that are not included in the training sets. As shown in Figs. 3(a) and 3(b), the band structures obtained from the predicted Hamiltonians are nearly indistinguishable from those obtained by diagonalizing the reconstructed reference Hamiltonians, particularly for states near the Fermi level. These results confirm that the reconstructed quasiparticle Hamiltonian is compatible with the locality and equivariance constraints built into the DeepH architecture, at least over the configurational range represented by the datasets.

We then perform a more challenging test of cross-scale transferability by applying the same models, trained exclusively on $3 \times 3 \times 3$ supercells, to randomly displaced $6 \times 6 \times 6$ supercells. The agreement between the predicted and reconstructed-reference band structures remains excellent, as shown in Figs. 3(c) and 3(d). For diamond, the mean absolute error over the 20 bands closest to the Fermi level is 5.9 meV. For GaP, the corresponding error over the 40 closest bands is 9.7 meV. The slightly larger error for GaP is physically plausible because GaP is a polar insulator. Atomic displacements in a polar crystal can generate long-range dipolar potentials through nonzero Born effective charges, introducing contributions that cannot be fully captured by a strictly local model. Nevertheless, the observed accuracy indicates that the short-range neural network model adopted here captures the dominant quasiparticle-level information for the configurations studied. These results demonstrate that the effective Hamiltonian retains sufficient nearsightedness to support transfer from small-sized training structures to substantially larger ones.

*Study of GW-level electron-phonon renormalization with DeepH-GW*. — The demonstrated cross-scale transfer capability of DeepH-*GW* opens the possibility of studying many-body phenomena in systems that are exceptionally expensive for direct *GW* calculations. Electron-phonon coupling provides an important example. Electron-phonon interactions are usually the major source of the temperature dependence of semiconductor band gaps, and previous studies have shown that many-body corrections are often required for quantitatively accurate predictions [53–55]. Direct finite-displacement calculations require repeated *GW* calculations in large supercells to sample phonon modes adequately and are thus highly expensive. To solve this issue, various strategies have been developed to reduce the size of supercells or reduce the number of required configurations [56–58]. Meanwhile, linear-response *GW*PT completely avoids explicit supercell calculations by working in the primitive cell [40,42]. Nevertheless, all these calculations still remain substantially more expensive than conventional *GW*. By enabling efficient predictions of quasiparticle Hamiltonians in large supercells with thermal displacements, DeepH-*GW* offers a complementary route to this problem.

We demonstrate this capability by calculating the electron-phonon renormalization of the diamond band gap using a finite-displacement approach [56,57] at the *GW* level. Density-functional perturbation theory calculations are first performed in the primitive cell with Quantum ESPRESSO [50] to obtain the phonon dispersion. The Zacharias-Giustino (ZG) special-displacement method [56,57] is then used to generate $12 \times 12 \times 12$ supercells with displacements. The ZG method provides an efficient approximation to configurational sampling from a finite-temperature ensemble. For each temperature, we generate ten configurations to reduce the randomness introduced by sampling. The effective AO *GW* Hamiltonian of each configuration is predicted using the model trained only on $3 \times 3 \times 3$ diamond supercells. The calculations are very cheap: neural-network inference requires approximately 5 CPU-hours per structure, while sparse diagonalization of the AO Hamiltonian requires approximately 14 CPU-hours per $\boldsymbol{k}$ point to obtain 128 eigenstates around the Fermi level. We should note that using the ZG method with the *GW* Hamiltonian captures the effects of electron-phonon interactions at the *GW* level on the electronic states. Following Refs. [56,57], the resulting supercell bands are unfolded into the primitive-cell Brillouin zone to obtain spectral functions. These spectral functions are then averaged over the sampled configurations, after which the temperature-dependent band energies are extracted.

Figure 4 shows the resulting temperature dependence of the fundamental band gap of diamond. The DFT-level calculation from Ref. [53], obtained using density-functional perturbation theory, underestimates the magnitude of the band-gap renormalization. In contrast, the DeepH-*GW* prediction is in substantially better agreement with experiment. This result demonstrates that DeepH-*GW* can preserve physically meaningful many-body corrections when transferred from relatively small training cells to thermally displaced supercells containing thousands of atoms.

## Discussion

We have introduced DeepH-$GW$, a framework for predicting effective $GW$ quasiparticle Hamiltonians directly from atomic structure. Three fundamental and complementary components are utilized and combined in this work, including high-fidelity PW $GW$ reference calculations, a real-space reconstruction framework that expresses the quasiparticle Hamiltonian in a localized AO subspace, and an E(3)-equivariant message-passing model that enables efficient prediction in large structures with high fidelity. In the future, with sufficiently broad and diverse training data, the DeepH-$GW$ approach could enable simulations of moiré materials, disordered systems, quantum defects, interfaces, and finite-temperature configurations containing thousands of atoms or even more. The learned quasiparticle Hamiltonians could also be connected with other advanced methods for excited-state phenomena such as optical response and nonequilibrium dynamics. By bypassing the historically most challenging and computationally intensive parts of $GW$ calculations, DeepH-$GW$ serves as a practical engine for large-scale simulation and materials discovery.

The present results also identify several areas for improvement. First, the reported neural-network errors are measured relative to reconstructed AO Hamiltonians and should be distinguished from the separate error introduced when PW quasiparticle information is projected into a specific AO subspace. Systematic optimization of the AO basis sets and band windows will therefore be important for improving the end-to-end accuracy. Second, broader benchmarks of the method should include chemically diverse materials, low-dimensional systems, metals, and so on. Systems containing slowly decaying electrostatic interactions or metallic responses may require explicit long-range corrections in addition to local message passing. Furthermore, several extensions of the reference $GW$ calculations are also desirable. The calculations presented here use one-shot $G_0W_0$ together with a generalized plasmon-pole model for the frequency dependence. Future work should examine full-frequency $GW$, spin-orbit coupling, off-diagonal self-energy matrix elements, DFT with different functionals as starting points, and partially or fully self-consistent $GW$ schemes. Additionally, incorporating DFT-level electronic descriptors could further improve accuracy and data efficiency.

## Acknowledgements

The work was supported by the U.S. National Science Foundation (NSF) under Grant No. DMR-2325410 for development of deep-learning electronic structure and projection techniques and Grant No. OAC-2513830 in the development of interoperable software and the special displacements method. This work was supported by the Center for Computational Study of Excited-State Phenomena in Energy Materials (C2SEPEM) at Lawrence Berkeley National Laboratory (LBNL), which is funded by the U.S. Department of Energy (DOE), Office of Science, Basic Energy Sciences, Materials Sciences and Engineering Division under Contract No. DEAC02-05CH11231, as part of the Computational Materials Sciences Program, which provided advanced codes and computation of $GW$ results. This work was supported by the National Key Basic Research and Development Program of China (grants no. 2024YFA1409100, and no. 2023YFA1406400), the Basic Science Center Project of NSFC (grant no. 52388201), the National Natural Science Foundation of China (grants no. 12334003, no. 12421004, and no. 12361141826), Fundamental and Interdisciplinary Disciplines Breakthrough Plan of the Ministry of Education of China (JYB2025XDXM408), the NSFC Basic Research Scheme for PhD Students (grant no. 124B2072), and the China Postdoctoral Science Foundation (grant no. 2025M773367). This research used computational resources provided by the National Energy Research Scientific

Computing Center (NERSC), a Department of Energy User Facility using NERSC award No. BES-ERCAP0036104, and the Frontera computing project at the Texas Advanced Computing Center (TACC) at The University of Texas at Austin, which is supported by National Science Foundation award No. OAC-1818253. This research used resources of both the Argonne and Oak Ridge Leadership Computing Facilities, which are DOE Office of Science User Facilities supported under contracts DE-AC02-06CH11357 and DE-AC05-00OR22725.

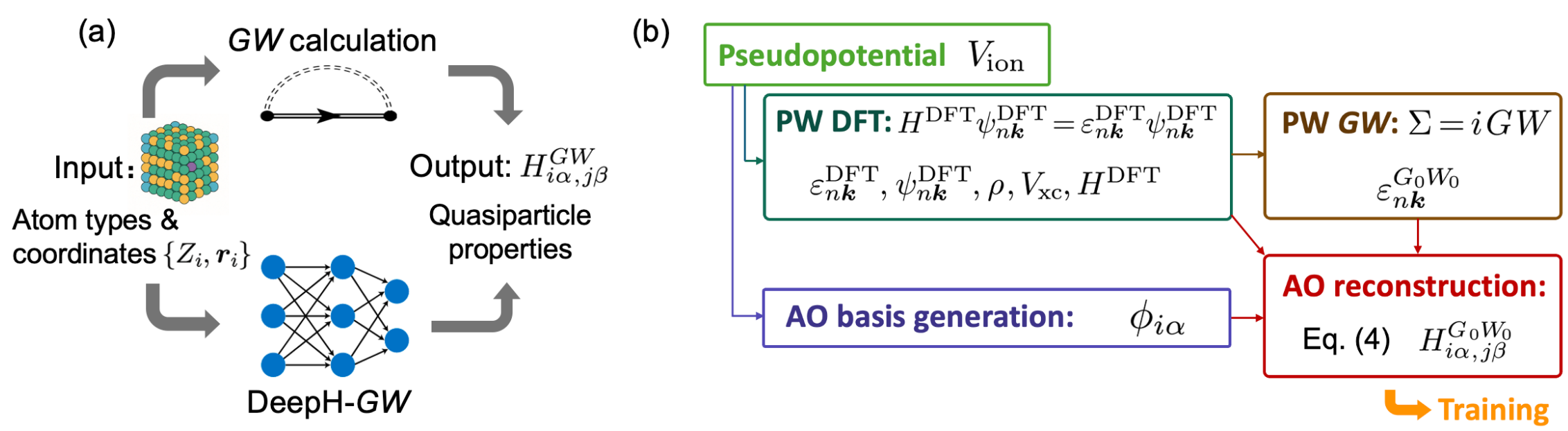

FIG. 1. Overview of the DeepH-*GW* method. (a) An effective *GW* Hamiltonian is constructed in a localized atomic-orbital (AO) basis from a first-principles high-fidelity *GW* calculation. Diagonalization of this Hamiltonian yields quasiparticle energies and wavefunctions from which excited-state properties can be evaluated. DeepH-*GW* learns the mapping from atomic structure to this reconstructed AO Hamiltonian. (b) Workflow for constructing the effective AO *GW* Hamiltonian according to Eq. (4). A plane-wave (PW) self-consistent DFT calculation provides the starting point for the PW *GW* calculation. In parallel, an AO basis is generated using an AO DFT calculation with the same set of pseudopotentials. The resulting DFT and *GW* quantities are then collected to construct the effective Hamiltonian used for model training.

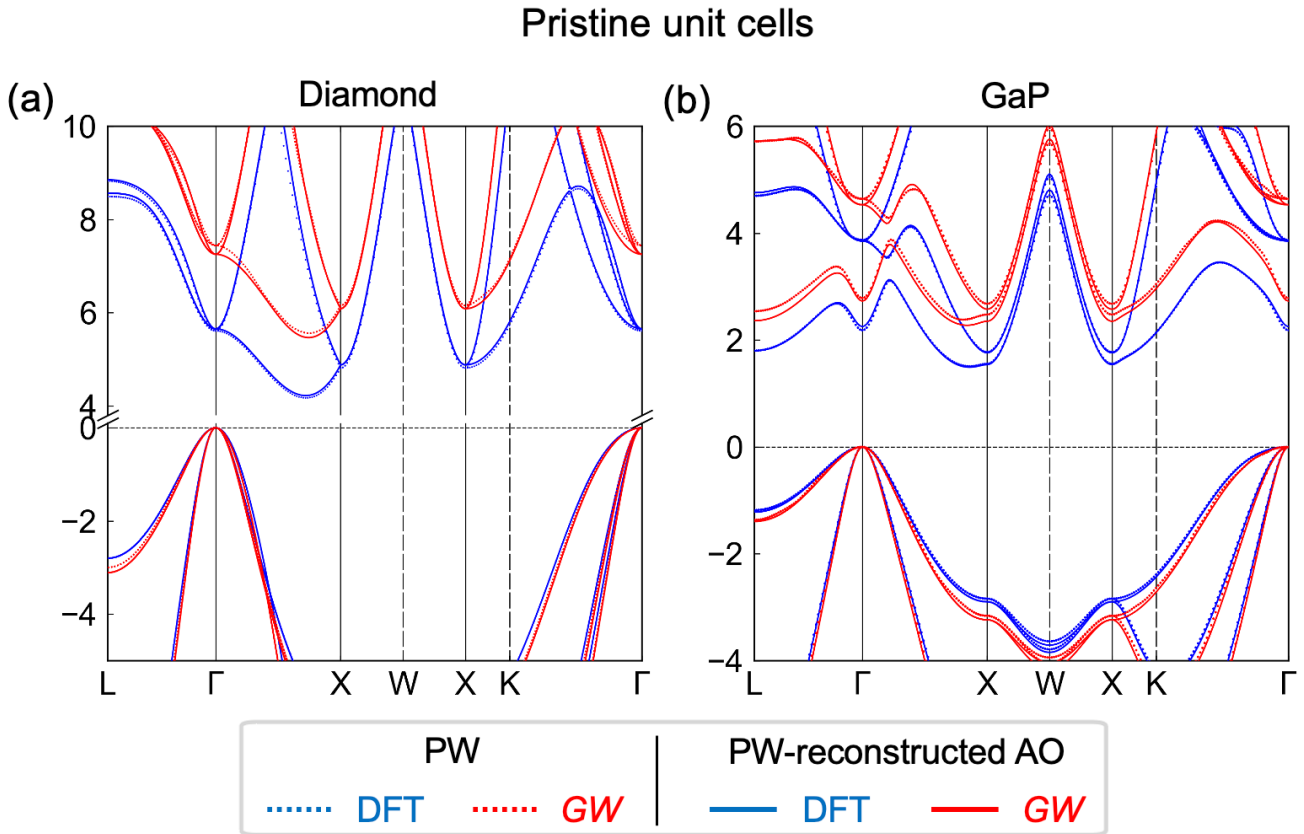

FIG. 2. Validation of the reconstructed AO Hamiltonians against PW calculations at the DFT and *GW* levels. At the DFT level, the band structures obtained from the reconstructed AO Hamiltonians (blue lines) agree closely with the PW DFT results (blue dots) for (a) diamond and (b) GaP. At the *GW* level, the reconstructed AO effective *GW* Hamiltonians (red lines) reproduce the overall quasiparticle band structures obtained from PW *GW* calculations (red dots), with discrepancies for some conduction bands. In each panel, the valence-band maximum is set to zero.

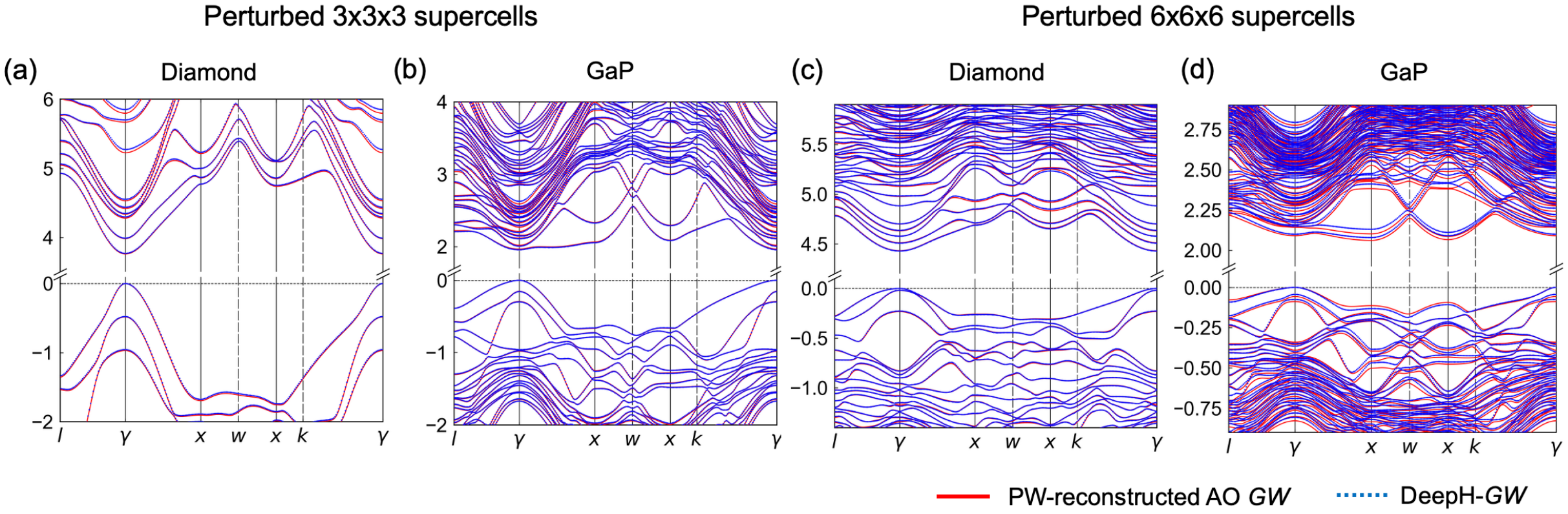

FIG. 3. Accuracy and cross-scale transferability of DeepH-$GW$. (a) Band structure predicted by a DeepH-$GW$ model trained on randomly displaced $3 \times 3 \times 3$ diamond supercells and evaluated on an unseen supercell of the same size. (b) Corresponding result for GaP. (c) Prediction for a randomly displaced $6 \times 6 \times 6$ diamond supercell using the model trained only on $3 \times 3 \times 3$ structures. The mean absolute error (MAE) over the 20 bands closest to the Fermi level is 5.9 meV. (d) Corresponding cross-scale prediction for GaP. The corresponding MAE over the 40 closest bands is 9.7 meV. In all panels, the DeepH-$GW$ predictions (blue dots) are compared with the eigenvalues of the reference PW-reconstructed AO effective $GW$ Hamiltonians (red lines).

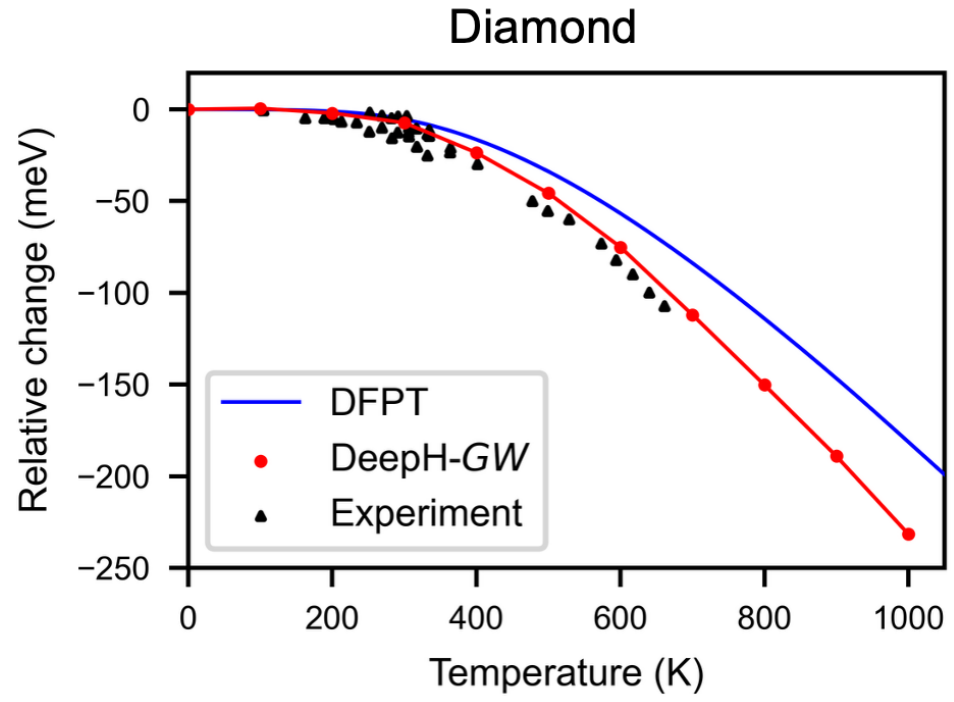

FIG. 4. Temperature dependence of the fundamental (indirect) band gap of diamond. The temperature-dependent fundamental band gap computed using DeepH-$GW$ is evaluated using the Zacharias-Giustino special-displacement method [56,57] averaged over an ensemble of ten $12 \times 12 \times 12$ supercells. The DFT-level calculations with electron-phonon matrix elements from DFPT (blue line) underestimate the magnitude of the band-gap renormalization, whereas direct DeepH-$GW$ simulations (red dots) yield substantially better agreement with experiment (black triangles). All values are reported relative to the corresponding $T = 0$ result. Thermal-expansion effects are included, and the red line serves as a guide to the eye. Experimental data are from Clark et al. [59].